\documentclass[epj,referee]{svjour}
\usepackage{latexsym}
\usepackage{graphics}
\usepackage{graphicx}
\usepackage{amssymb}
\usepackage{epstopdf}
\begin{document}
\title{Statistical mechanical approach to secondary processes and structural 
relaxation in glasses and glass formers}

\subtitle{A leading model to describe the onset of  Johari-Goldstein processes 
and their relationship with fully cooperative processes}
\author{Andrea Crisanti\inst{1,2}\and Luca Leuzzi\inst{2,3}
\and Matteo Paoluzzi\inst{3,4}}

\institute{Dipartimento di Fisica, Universit\`a {\em Sapienza}, Piazzale Aldo Moro, 5 - 00185 - Rome, Italy \and CNR-ISC, Via dei Taurini, 19 - 00185 - Rome, Italy \and CNR-IPCF, UOS Roma {\em Kerberos}, Piazzale Aldo Moro, 5 - 00185 - Rome, Italy \and Dipartimento di Fisica, Universit\`a Roma Tre, Via della Vasca Navale, 84 - 00184 - Rome, Italy}
\date{Received: date / Revised version: date}
\authorrunning{A. Crisanti, L. Leuzzi, M. Paoluzzi}
\titlerunning{Statistical mechanical approach to secondary processes}
\abstract{
The interrelation of dynamic processes active on separated
time-scales in glasses and viscous liquids is investigated using a model displaying two time-scale bifurcations
both between fast and secondary relaxation and between secondary and structural relaxation. The study of the dynamics allows for predictions on
the system relaxation above the temperature of dynamic arrest in the mean-field approximation, that are compared with the outcomes of the equations of motion directly
derived within the Mode Coupling Theory (MCT) for under-cooled viscous
liquids. Varying  the external thermodynamic parameters a wide range of phenomenology can be represented, from a very clear separation of structural and secondary peak in the susceptibility loss to excess wing structures.
} %end of abstract
\maketitle
\section{Introduction}
\label{intro}

Secondary processes in supercooled liquids and glasses are related  to complicated though local, non- or not fully cooperative, dynamics. They occur on time-scales much slower than cage rattling, but much faster than structural relaxation. Their existence was first pointed out in the 1960's from the experimental observation of a second peak in dielectric loss spectra at a frequency, $\nu_\beta\sim 1/\tau_\beta$, higher than the frequency $\nu_\alpha\sim 1/\tau_\alpha$ of peak known to represent the structural $\alpha$ relaxation.
This so-called  $\beta$-peak was recorded in glycerol, propyleneglycol, n-propane, different polymeric substances and  liquids composed of rigid molecules. Johari and Goldstein eventually conjectured that such processes - now known as Johari-Goldstein (JG) - originate from  the same complicated frustrated interactions leading to the glass transition  \cite{Johari70,Johari70b,Johari71}.

Also in cases where spectral density of response losses do not clearly show a second peak, secondary processes can be active and induce some anomaly at high frequency. This feature of the susceptibility loss part  is called "excess wing" and was initially observed as an apart phenomenon \cite{Wong74}.
Actually, classifications exist in terms of glass formers displaying excess wings and substances showing well defined $\beta$-peaks \cite{Adichtchev03,Blochowicz03,Adichtchev03b}.
Although more recent investigation has provided evidence supporting the idea that
the excess wing is not an apart dynamic process, but rather a manifestation of a JG process
\cite{Ngai01}\cite{Ngai04b} and that tuning proper thermodynamic parameters (temperature, pressure, concentration, ...) the latter can emerge out of the first one (or, viceversa, a secondary peak can reduce to an excess wing). Cummins \cite{Cummins05} suggests, e.g.,  that the relevant parameter may be the rotation - translation coupling constant which becomes stronger as density increases, because of pressure increase or temperature decrease, and is larger for liquid glass former made of elongated, strongly anisotropic molecules. Also theoretical attempts have been carried out in this direction as, for instance, in the framework of Mode Coupling Theory (MCT), by means of which the relaxation of reorientational correlation and rotation-translation coupling in liquids composed of strongly anisotropic molecules appears to be logarithmic in time \cite{Goetze04}.

A comprehensive picture is, though, not yet established  and many questions are open. For instance about the dependence on temperature and pressure (or concentration) of characteristic time scales of JG processes, or the possibility that seconday processes might disclose a certain degree of cooperativeness \cite{Stevenson10}, or the persistence of $\beta$ processes also below the glass transition temperature $T_g$ \cite{Cummins05}.  A very interesting question is if there is a straightforward connection, and, in case, which one, between processes evolving at qualitatively different time-scales. Or, rephrased, whether one might devise the long-time behavior of $\alpha$ relaxation from the behaviors of the fast small amplitude cage dynamics ($\gamma$ processes) and of the (slower) secondary processes.

In glasses, and glass-formers, where $\alpha$ and JG $\beta$ peaks can be clearly resolved in frequency (e.g., 4-polybutadiene, toluene\cite{Wiedersich99} \cite{Kudlik99} or sorbitol \cite{Nozaki98}), one can describe the system in terms of a scenario  where two time-scale bifurcations accelerate as temperature is lowered and processes consequently evolve on three "well separated" time sectors.

We, therefore, analyze the dynamic properties of a model for slowly relaxing glassy systems with up to three time-scales.
This is a generalization of the $p$-spin model with quenched disorder, that is known to heuristically reproduce  all the basic features of structural glasses \cite{Kirkpatrick87b} \cite{Kirkpatrick89} \cite{Crisanti92} \cite{Crisanti93} and whose dynamics above a certain temperature ("dynamic" or "mode coupling") is equivalent to the dynamics of the schematic mode coupling theory (MCT) with a kernel depending from the correlator as $\phi^{p-1}$  \cite{Bouchaud96}. The generalization consists in coupling the dynamical variable ÒspinÓ (playing, e.g.,  the role of a density fluctuation, or a component of molecular orientation)
with other spins in two different ways: 
as a part of a group of $s$ variables and as a part of a group of $p$  variables. Variables in each group interacts among themselves  through i.i.d. random multi-body interaction of zero mean and mean square strength of magnitude $\sim J_s$ and $J_p$, respectively.
As one of these two interaction mechanisms (e.g., "$p$") involves sensitively more dynamical variables than the other (e.g., "$s$"), this triggers a mixture of strong and weak  cooperativeness that can be varied by an external control parameter (e.g, $J_{p}/J_s$).

Our aim is to provide a model to interpolate between different resolutions of secondary processes and support the idea that excess wings and secondary peaks are both manifestations of  ÒintermediateÓ (slow, yet thermalized) processes between cage rattling and structural relaxation.
Relying on the results about correlation functions and spectral densities we can argue on the possible interrelation between processes evolving on different time-scales and their characteristic times, $\tau_{\alpha}$, $\tau_\beta$ and $\tau_{\gamma}$.

\section{The leading spin model for secondary processes}
\label{sec:model}

The model we will consider is a
spherical $s+p$-spin interaction  model:
\begin{equation}
\label{f:Ham}
{\cal H} = \! \sum_{i_1<\ldots <i_s}\!J^{(s)}_{i_1\ldots i_{s}}
\sigma_{i_1}\cdots\sigma_{i_s}
       +\!\sum_{i_1<\ldots <i_p}\!J^{(p)}_{i_1\ldots i_p}
       \sigma_{i_1}\cdots\sigma_{i_p}
\end{equation}
where $ J^{(t)}_{i_1\ldots i_{t}}$ ($t=s,p$) are uncorrelated, zero
mean, Gaussian variables of variance $J_t^2 t!/(2N^{t-1})$ and
$\sigma_i$ are $N$ ``spherical spins'' obeying the constraint $\sum_i
\sigma_i^2 = N$.
Since every spin interact (very slightly, $J_t\sim 1/{N}^{(t-1)/2}$) with every other, for this system the mean-field approximation is exact.
We will consider the case in which each spin interact with the rest of the system in two different ways: in {\em small} groups (of $s$ elements)
and in {\em large} group (of $p$ elements). If $p-s$ is large enough
standard MCT provides evidence for glass-to-glass transitions beyond the line of validity of time translational invariance \cite{Krakoviack07}, that is a fundamental assumption for  MCT.
The theories developed for  quenched disordered systems, allow further to compute the stable solutions corresponding to the glassy phases involved below the dynamic transition and identify the nature of the processes ongoing in each one of the glasses. Eventually, it can be shown that  the model thermodynamics displays three distinct glass phases below the line of dynamic arrest, one of which consisting of processes thermalized at three completely separate time-scales \cite{Crisanti07}\cite{Crisanti07b}.
Starting from these considerations dynamic equations are obtained, reducing to those of schematic MCT above the mode coupling temperature $T_{d}$. We will see that the three time-scale glass  will be already signaled  in the dynamics following educated paths in the phase diagram.

The glass phase with double bifurcation of time-scales can be yielded in the $s$-$p$ spherical spin model under a certain condition on the values of $s$ and $p$, i.e., for a given $s$, $p$ must be equal or larger than value solution of 
\begin{equation}
(p^2+s^2+p+s-3ps)^2-ps(p-2)(s-2)=0.
\end{equation}
as it has been shown in Ref. \cite{Crisanti07}.
Some example of "threshold" couples $(s,p)$
to obtain a double bifurcation are $(3,8)$, $(4,11)$ or $(5,16)$.
The larger is $p-s$, the broader the region of phase diagram where double bifurcation can be found.

The external thermodynamic parameters are the 
temperature and the relative weight of the 
two interaction terms (big to small) in the Hamiltonian. In unit of $J_s$:  $T/J_s$ and  $J_p/J_s$. 
These are related to the usual mode-coupling parameters  so that the memory kernel of the dynamic equation takes the mode-coupling form 
\begin{eqnarray}
{\cal K}(\phi)&=& \mu_s \phi^{s-1}+\mu_p\phi^{p-1}
\label{f:kappa}
\\
\mu_p&=&p\beta^2 J_p^2/2
\\
\mu_s&=&s\beta^2J_s^2/2
\end{eqnarray}

We stress that as $p-s$ is large, and $s>2$, the theory we are considering yields qualitatively different results from schematic MCTs with, e.g., $s=2$ and $p=3$ \cite{Goetze04,Greenall07}. Indeed, in schematic MCT with linear and quadratic terms in the kernel a clear separation of time scales is unfeasible and the (possible) thermodynamic glassy phase underneath can only provide an acceleration of time-scales bifurcation, as discussed in Refs. \cite{Crisanti04b,Crisanti06,Krakoviack07,Crisanti07b}.

The strong three-level separation we study can, then, be softened and adapted to less defined structures than the two peaks  (e.g., the excess wing), by tuning the external parameters temperature and $J_p/J_s$, or by choosing  $s,p$ model instances with  smaller $p-s$.

The model was initially developed to study the nature of polyamorphism and
amorphous-to-amorphous transitions.  On the static front, the analysis can be carried out within the framework of Replica Symmetry Breaking (RSB) theory, leading to the identification of low
temperature glass phases of different kinds \cite{Crisanti07}.
Below the Kauzmann-like transition line $T(J_{p}/J_s)$, the model displays both "one-step" RSB solutions, known to reproduce all basic properties of structural glasses \cite{Kirkpatrick89}, and a physically consistent "two-step" solution \cite{Crisanti07}.

Above the Kauzmann transition line, the thermodynamic stable phase is the fluid paramagnetic phase but excited glassy metastable states are present in a large number, growing exponentially with the size $N$ of the system. The configurational entropy of the system is, thus, extensive.
Because barriers between minima of the free energy landscape separating local glassy minima  grow like some positive power of $N$ in the mean-field approximation, "metastable" states have, actually, an infinite life-time in the thermodynamic limit and ergodicity breaking occurs as soon as an extensive configurational entropy appears. The highest temperature at which this happens is known as {\em dynamic} \cite{Crisanti93}, {\em arrest} \cite{Kirkpatrick89} or {\em Mode Coupling} \cite{Goetze09} temperature. We shall denote it by $T_d$. As the temperature is lowered down to $T_d$ the spin-spin time correlation function (analogue of the correlation between density fluctuations) develops a plateau that, eventually, extends to infinite time as $T=T_d$, signaling the breaking of the ergodicity.

\begin{figure*}[t!]
\center
\includegraphics{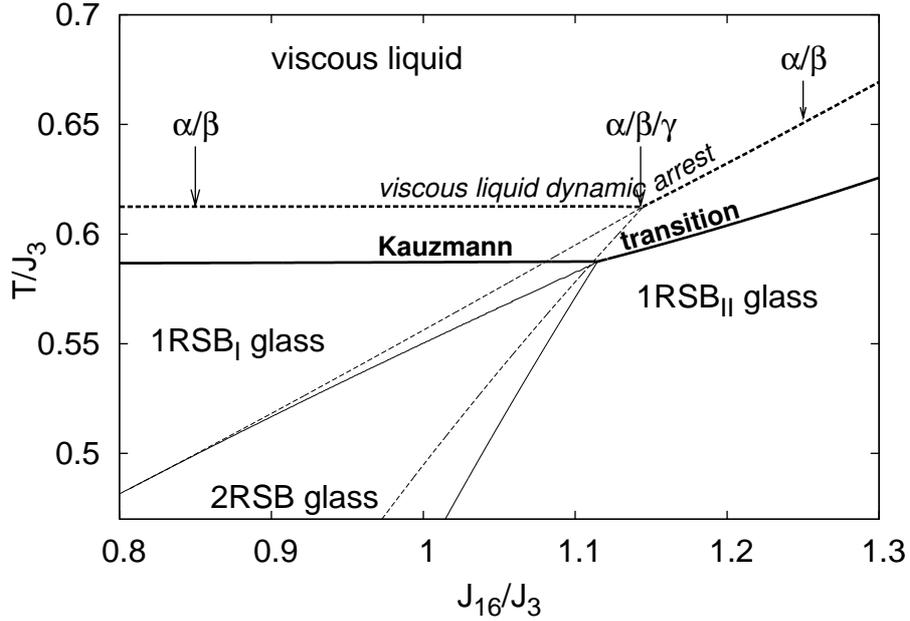}
%\vspace*{.1cm}       
\caption{Phase diagram of the $s=3$, $p=16$ spherical spin model.  Dynanic transition lines are dashed and thermodynamic transition lines beneath are full. 1RSB$_I$ glass stays for a glass with a single time scale bifurcation with relatively low nonergodicity factor for the time correlation function. 1RSB$_{II}$ glass stays for a glass with a single time scale bifurcation with higher  nonergodicity factor. The 2RSB glass displays two bifurcations and two possible correlation values in the arrested state.}
\label{fig:phdi}
\end{figure*}

In Fig. \ref{fig:phdi} we display the $(T/J_s, J_p/J_s)$ phase diagram for $s=3$ and $p=16$. We will use this specific case throughout the paper, for which strong discrimination of the secondary
processes is easily realizable in a relative wide region of the phase diagram. The dynamic and thermodynamic properties of such an instance below the dynamic transition are discussed in Ref. \cite{Leuzzi08b}.

Since the dynamic counterpart of a RSB is known to be a time-scale bifurcation \cite{Sompolinsky81,Sompolinsky82}, Eq. (\ref{f:Ham}) provides a leading
model to probe the behavior of characteristic time-scales in presence of secondary processes and the different mechanisms in which they can arise starting from high temperature and cooling down the system.  

\subsection{Dynamics}
\label{sec:dynamics}

The relaxational dynamic of the system is described by the Langevin equation
\begin{eqnarray}
\Gamma_0^{-1}\frac{\partial\sigma_k(t)}{\partial t}&=& -\frac{\delta {\cal H}[\{\sigma\}]}{\delta \sigma_k(t)}+\eta_k(t)
\\
\nonumber
&&\langle \eta_k(t)\eta_n(t')\rangle=2 k_B T \Gamma_0^{-1} \delta_{kn}\delta(t-t')
\end{eqnarray}
where $\eta_k$ is the thermal white noise and $\Gamma_0^{-1}$ is the microscopic time-scale.
Using a Martin-Siggia-Rose path-integral formalism one can reduce the equations of motion to a single variable ($\sigma(t)$) formulation \cite{Martin73,Dedominicis80}.
The fundamental observables to study the onset of a slowing down of the dynamics are the time correlation of between the spin variable at time $t'$ and time $t>t'$ and the response function to a small perturbative field $h$. For our
system they are defined as
\begin{eqnarray}
C(t,t') &=& {\overline {\langle \sigma(t)\sigma(t') \rangle}}
\\
G(t,t')&= &\frac{\delta\overline{\langle \sigma(t)\rangle}}{\delta \beta h(t')}; 
 \qquad t>t'
\end{eqnarray}
where the overbar denotes the average over quenched disorder, wheres the brackets stay for an average over different trajectories (thermal average).
For temperature above $T_d$ the time translational invariance (TTI) holds and  the response and correlation functions are related by the Fluctuation - Dissipation Theorem (FDT):
\begin{equation}
G(t-t')=\theta(t-t')\partial_{t'}C(t-t')
\end{equation}
The dynamic equation of the correlation function then takes the form
\begin{equation}
\Gamma_0^{-1}\frac{\partial C(t)}{\partial t}+\bar r C(t)+\int_0^t d t'\Lambda[C(t-t')]\frac{ \partial C(t')}{\partial t'} = \bar r -1
\label{eq:dyn}
\end{equation}
with initial condition $C(t=0)=1$, and
\begin{equation}
\bar r = r-\Lambda[C(t=0)]
\end{equation}
The parameter $r$ is the "bare mass" \cite{Crisanti08}, that for the spherical model is related to the Lagrange multiplier used to impose the spherical constraint \cite{Crisanti93}. The value of $\bar r$ depends on temperature, and $J_{s,p}$; however, in the high temperature phase it is constant and equal to $1$, so that the r.h.s. of (\ref{eq:dyn}) vanishes.

The function $\Lambda(t)=\Lambda[C(t)]$ is the memory kernel that in the specific case of our model has the functional form:
\begin{equation}
\Lambda(q) = \mu_s q^{s-1}+\mu_p q^{p-1}
\label{def:Lambda}
\end{equation}
to be compared with Eq. (\ref{f:kappa}).
Indeed, the evolution of the correlation function is described by a dynamical equation equivalent to  that of schematic mode-coupling  theories, i.e., in which the second order time derivative term in MC equations 
%is neglected \cite{Bouchaud96,Goetze09}.
is replaced by the first order one \cite{Bouchaud96,Goetze09}.

\begin{figure}
\resizebox{0.95\columnwidth}{!}{%
\includegraphics[angle=270]{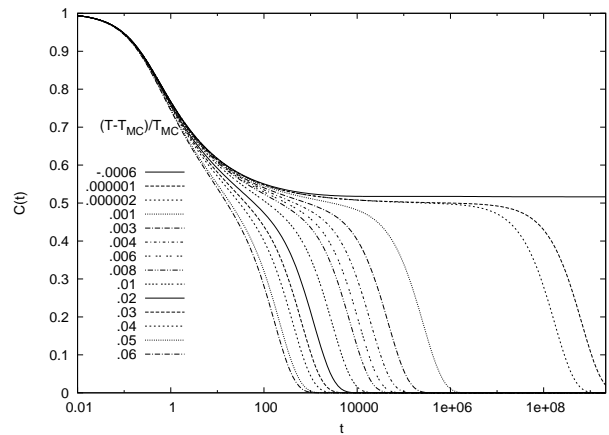}}
\caption{Correlation function vs. time on log-scale at fixed temperature for the $s=3$-spin model, i.e., $J_p=0$. }
\label{fig:J_16_0}       % Give a unique label
\end{figure}

For $J_s=0$ one recovers the usual spherical $p$-spin model \cite{Crisanti93}. In such model, above  $T_d$ the correlation function has the shape plotted in Fig.
\ref{fig:J_16_0}, with one plateau developing for long time.

\begin{figure}
\resizebox{0.95\columnwidth}{!}{%
\includegraphics[angle=270]{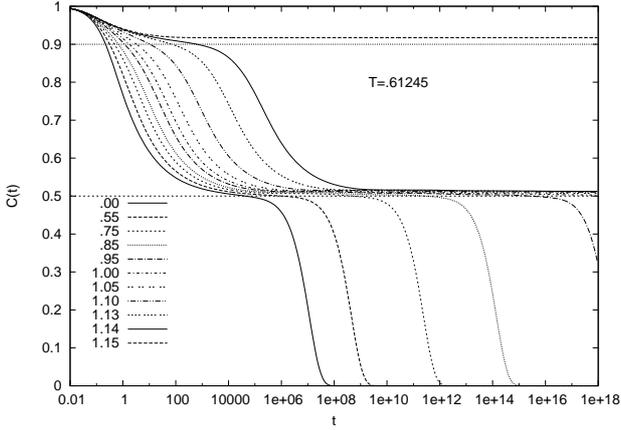}}
\caption{ Correlation function vs. time on log-scale at fixed $T/J_s=0.61245$
with $s=3$ and $p=16$ increasing  $J_{16}$ from zero to $J_{16}/J_3= 1.145$ such that $T_d(J_{16}/J_3)= 0.61245$.}
\label{fig:J_16_Tfix}
\end{figure}

\begin{figure}
\resizebox{0.95\columnwidth}{!}{%
\includegraphics[angle=270,width=0.95\columnwidth]{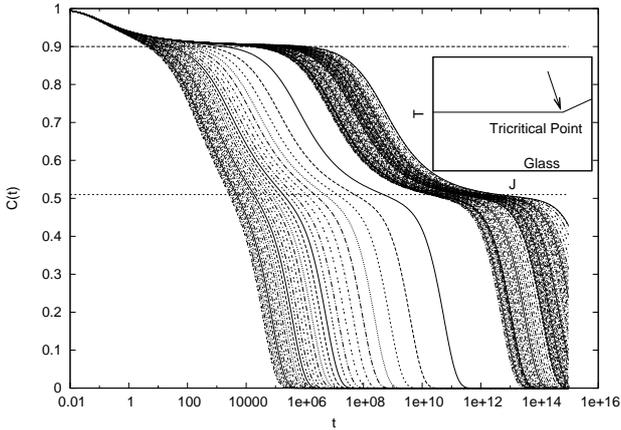}}
\caption{Correlation function vs. time on log-scale in a cooling procedure in the $T/J_s,J_p/J_s$ phase diagram with $s=3$ and $p=16$ along a path perpendicular to the right hand side fluid/glass dynamic transition line,
ending at the tricritical point ($0.61234,1.1446$). }
\label{fig:T3crit_path}
\end{figure}

\begin{figure}
\resizebox{0.95\columnwidth}{!}{%
\includegraphics[angle=270,width=0.95\columnwidth]{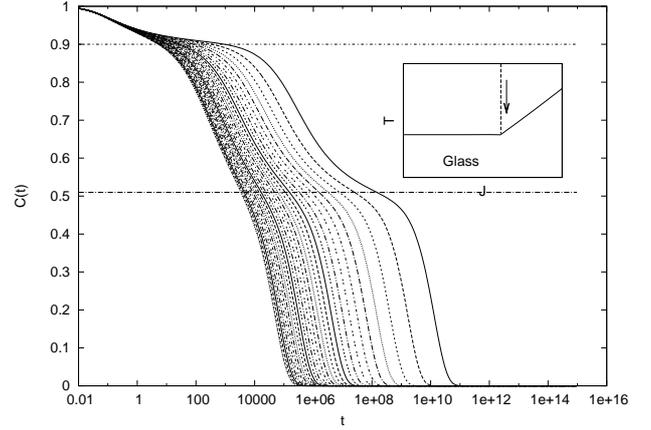}}
\caption{Correlation function vs. time on log-scale in a cooling procedure in the $T/J_s,J_p/J_s$ phase diagram with $s=3$ and $p=16$ along the constant $J_{16}=0.1446$ line, i.e. path perpendicular to the left hand side fluid/glass dynamic transition line ending at the tricritical point. }
\label{fig:T3crit_path2}       % Give a unique label
\end{figure}

Cooling down the system and increasing the $J_s$ along certain paths in the phase diagram in order to approach the tricritical point, the time-correlation function develops {\em two} plateaus at different correlation values, cf. Figs.  \ref{fig:J_16_Tfix}, \ref{fig:T3crit_path} and \ref{fig:T3crit_path2}.

%%%%

As mentioned above we will denote by $\gamma$ the fastest relaxation (also referred to as $\beta_{\rm fast}$ \cite{Donth01}), by $\beta$ the secondary Johari-Goldstein relaxation
($\beta_{JG}$) and by $\alpha$ the structural relaxation.
In Fig. \ref{fig:T3crit_path} we display the behavior of $C(t)$ approaching from high temperature the tricritical point along a $T(J_p)$ line perpendicular to the dynamic transition line with the 1RSB$_{II}$ glass. Changing path, cf. Fig. \ref{fig:T3crit_path2} the qualitative behavior is the same (though  quantitative differences can be non-negligible).
A first plateau, $q_1$, occurs for $t\gtrsim t_\gamma$ and a second one, $q_2<q_1$, on the characteristic time-scale at which the  secondary relaxation occurs ($t\gtrsim t_\beta$). We, thus, study the behavior in $T$ of the characteristic relaxation times for processes on
different time-scales and their functional interrelation.

%%%
%%%%

%

Near each plateau $q_k$ the dynamical equation (\ref{eq:dyn}) predicts a power law behaviour of $C(t)$, with
\begin{equation}
C(t) -  q_\kappa \sim   t^{-a_\kappa},
\label{eq:a_k}
\end{equation}
for $C(t) \gtrsim q_k$, and the von Schweidler law:
\begin{equation}
C(t) - q_\kappa \sim -t^{b_\kappa}
\label{eq:b_k}
\end{equation}
for $C(t) \lesssim q_\kappa$.
We can now expand the dynamical equation  (\ref{eq:dyn}) about the plateaus in powers of
$\phi(t)= C(t)-q_\kappa$, with $\phi \ll 1$. To this aim, a suitable rescaled time $\tau = t / t_\kappa$, is introduced, where $t_{\kappa}$ diverges at the critical point, and a relative rescaling function $g_\kappa(\tau)$, such that $\phi(t)\sim g_\kappa(\tau) \sqrt{\bar r(q)-\bar r}$, cf. App. \ref{appendix1}.
Eventually  one obtains the scaling equation
\begin{equation}
%-\overline{m}_\kappa g^2_\kappa(t)+\frac{\partial }{\partial \tau_k} \int_0^\tau %d\tau'g_\kappa(\tau_\kappa-\tau')g_\kappa(\tau')=-1.
(1-\overline{m}_\kappa) g^2_\kappa(\tau)+\int _0^\tau d\tau'
\left[g_\kappa(\tau-\tau')-g_\kappa(\tau)\right]
\frac{\partial g_\kappa(\tauÕ)}{\partial \tauÕ}=-1.
\label{eq:rescaled}
\end{equation}
The parameter $\overline{m}_k$, also called "exponent parameter" $\lambda$ in MCT,
takes the exact expression
\begin{equation}
\overline{m}_\kappa = \frac{(1-q_\kappa)^3}{2}\Lambda''(q_\kappa)
\label{eq:exact_lambda}
\end{equation}
where the plateau correlations $q_k$ are obtained from the self-consistency equations for the asymptotic dynamic  solution for the 2RSB glass \cite{Leuzzi08b}.
Inserting the expressions (\ref{eq:a_k})-(\ref{eq:b_k}) of $g_k(t)$  into  Eq. (\ref{eq:rescaled}) one obtains the following relationships:
\begin{equation}
\overline{m}_\kappa=
\frac{\Gamma^2(1-a_\kappa)}
{\Gamma(1-2a_\kappa)};
\quad 0<a_\kappa<1/2
\end{equation}
and
\begin{equation}
\overline{m}_\kappa=\frac{\Gamma^2(1+b_\kappa)}{\Gamma(1+2b_\kappa)}; \quad 0<b_\kappa<1
\end{equation}
The analysis of the exponents for the two plateaus as the tricritical point is approached along the path perpendicular to the high $J_{16}$ dynamic transition line is written
in table \ref{tab:1} where we report their values for both plateaus.
The approach to the tricritical point is not unique and the estimate of the exponents is usually very sensitive in MCT. This can be a possible cause for the mismatch between numerically interpolated and 
theoretically computed, cf., Eq. (\ref{eq:exact_lambda})
\begin{table}
\center
\caption{Mode coupling theory exponents of power-law relaxation to and from high and low plateau in correlation.}
\label{tab:1}       % Give a unique label
\begin{tabular}{|ccc|ccc|}
\hline\noalign{\smallskip}
$a_1$& $b_1$ & ${\bar m_1}$ & $a_1$ (th) & $b_1$ (th) & ${\bar m_1}$ (th) \\
\noalign{\smallskip}\hline\noalign{\smallskip}
 0.38(1) & 0.89(1)& 0.54(1) & 0.38797& 0.95045 & 0.5252\\
\noalign{\smallskip}\hline
$a_2$ & $b_2$  &${\bar m_2}$ & $a_2$ (th) & $b_2$ (th)&${\bar m_2}$ (th)\\
\noalign{\smallskip}\hline\noalign{\smallskip}
0.302(3) & 0.55(1)  & 0.754 (6)& 0.30441& 0.55738 &0.7505\\
%number & number & number &\\
\noalign{\smallskip}\hline
\end{tabular}
% Or use
\vspace*{.2cm}  % with the correct table height
\end{table}

Moving to the frequency domain, the susceptibility loss, linked to the spectral densities by the Fluctuation - Dissipation Theorem
$\chi''(\omega)=\omega/(2T) S(\omega)$
 nearby the tricritical point displays two peaks as in the dielectric loss data of materials in which JG processes have been detected, cf., e.g.,  \cite{Nozaki98,Wiedersich99,Kudlik99,Ngai03}. 
 
 %In Fig. 
 %\ref{fig:J_16_zero}
% we plot the loss function at different temperatures along  the $J_p=0$ line. This corresponds to the $s$-spin model, in which no secondary peaks are expected.
 %The loss curves, derived from correlation functions with a single plateau,  are interpolated 
 %with the function
 %\begin{equation}
 %f(\omega)=\frac{\sin\left(\beta_{\rm cd}\arctan \omega\tau\right)}
 %{\left(1+\omega^2\tau^2\right)^{\beta_{\rm cd}/2}}
 %\end{equation}
 %i.e.,
 %the imaginary part of the Cole-Davidson function, frequency-domain analogue of the Kohlrausch-%Williams-Watts function for stretched exponential relaxation \cite{Lindsey80}.
 
\begin{figure}
\resizebox{0.95\columnwidth}{!}{%
\includegraphics[angle=270]{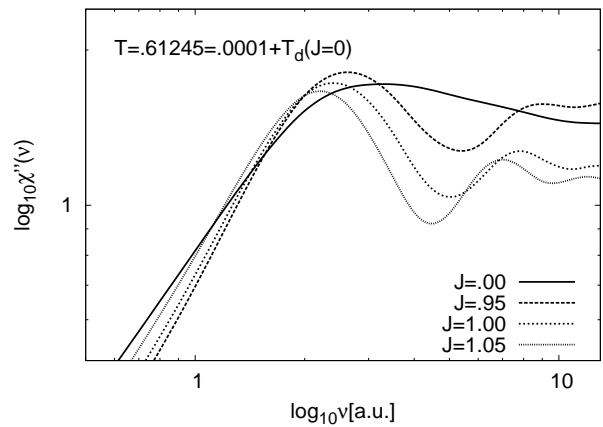}
}
\caption{Susceptibility loss in frequency $\omega$ at constant temperature $T=0.0001+T_d(J_{16}=0)$ and different values of $J_{16}/J_3$.}
\label{fig:J_16_Tfix_freq}       % Give a unique label
\end{figure}

 In Fig. \ref{fig:J_16_Tfix_freq} the development of the secondary peak is plotted as the tricritical point is approached in the $T,J_p$ diagram.
For small contribution from the $p$ interaction only the $\alpha$ peak is pronounced near the transition of dynamic arrest.
As the $p$-body interaction increases in strength and the tricritical point is approached  a secondary $\beta$ peak arises.

%

%%%%%%%%%

\section{Interrelation between relaxation times}
From the times at which the correlation decays from each -well separated- plateau we can investigate
the possibility of a functional relationship among them.
In Ngai's Coupling Model \cite{Ngai03,Ngai04}.
the evidence of a deep link between secondary and structural processes is, e.g.,  connected to a strong stretch in the exponential relaxation to equilibrium
in supercooled liquids   \cite{Kohlrausch47}\cite{Williams70}
\begin{equation}
C_{KWW}(t)=\exp\left\{-\left(\frac{t}{\tau}\right)^{1-n}\right\}
\end{equation}
by the law
\begin{equation}
\tau_\alpha = [t_c^{-n}\tau_\beta]^{1/(1-n)}; \quad 0<n<1,
\label{KLNlaw}
\end{equation}
with $t_c$ the time at which fast Maxwell-Debye exponential relaxation matches KWW relaxation.  
The larger is $n$, the more the peak at high frequency (short times) is pronounced. As $n$ is small no peak related to secondary
processes is appreciated.

\begin{figure}
\resizebox{0.95\columnwidth}{!}{%\center
\includegraphics[angle=270]{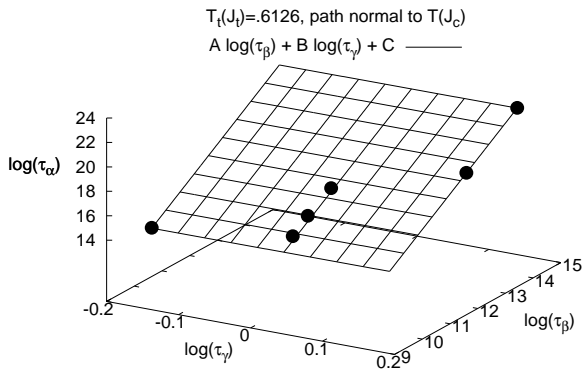}}
%\vspace*{.1cm}       
\caption{Interrelation between the characteristic relaxation times of the fast ($\gamma$), Johari-Goldstein ($\beta$) and fully cooperative ($\alpha$)
processes.}
\label{fig:taus}
\end{figure}
%%%%
\begin{figure}
\resizebox{0.95\columnwidth}{!}{%\center
\includegraphics{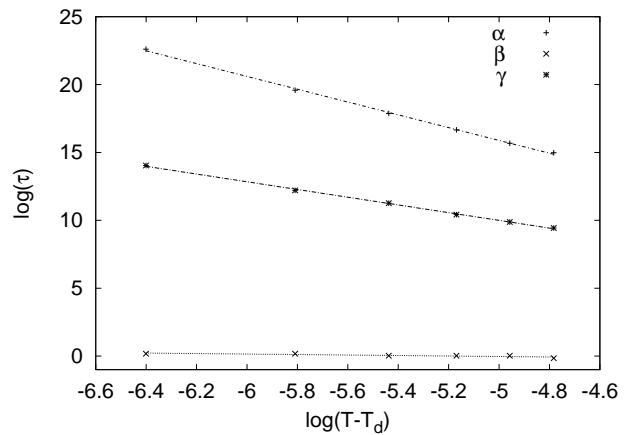}}
%\vspace*{.1cm}       
\caption{Behavior of $\tau_{\alpha}$, $\tau_\beta$ and $\tau_\gamma$ vs $T-T_d^{(3c)}$ along the phase diagram path perpendicular to the fluid/1RSB$_{II}$ dynamic transition line approaching the tricritical point.}
\label{fig:taus2}
\end{figure}

In our model the structural relaxation to equilibrium turns out to be purely exponential also very near  the dynamic transition temperature.
However, the relaxation at time scales larger than the $\tau_\beta$ (decay from highest plateau) does
present a non-exponential behavior containing, on top of the final fully cooperative relaxation at $\tau_\alpha$, also the relaxation to the lowest plateau (where $\beta$ processes are thermalized and $\alpha$ are completely stuck) and the decay from it, that follows the von Schweidler law, cf. Eq. (\ref{eq:b_k}).
\footnote{In MCT, it is, actually, common that stretched exponential relaxation only occurs at high wave-numbers. In our model we do not implement the wave-number dependence (we operate in the long distance limit).}
In this respect, the stretched  exponential might still be recovered and considered as an uneducated guess for the actual multi-time-scales dynamics. An alternative estimate of $n$ would then support such conjecture.
As a matter of fact, the relationship between fast, secondary and structural processes
appears to qualitatively follow Ngai's law, Eq. (\ref{KLNlaw})
in a generic form:
\begin{equation}
\log \tau_\alpha = \beta_0 \log\tau_\beta + \gamma_0 \log \tau_\gamma
\label{KLNgen}
\end{equation}
In Fig. \ref{fig:taus} we plot the inter-dependence of the relaxation times of separated processes and
the dependence on $\log \tau_\alpha$ on $\log \tau_\beta$ and $\log \tau_\gamma$ turns out to lie on a plane (with $\log\tau_\gamma$ almost constant ), confirming Eq. (\ref{KLNgen}).

In our description the time-scales over which one kind of process is active are well defined by characteristic times of relaxation to the plateaus. Cage rattling dynamics is thermalized already at the higher plateau of the correlation function and its equilibration time $\tau_\gamma$ does not depend on the distance from the dynamic critical point, cf. Fig. \ref{fig:taus2}.

Slower JG processes (of intermolecular origin) \cite{Johari70}\cite{Johari73} take place when structural relaxation is completely stuck and are strongly correlated off-equilibrium  for a time such that $C(t)\simeq q_2$. Their characteristic time grows several order of magnitude, yet remaining several order of magnitudes smaller than $\tau_\alpha$, cf., Fig.  \ref{fig:taus2}.
After that they relax to equilibrium on the characteristic time $\tau_\beta$ and the total correlation decreases to a second plateau $q_1$ where
the longest processes, the cooperative $\alpha$ processes, remain off-equilibrium  until $C(t)\simeq q_1$.  Eventually, structural relaxation goes towards equilibrium, on the characteristic time-scale $\tau_\alpha$.

\section{Conclusions}
We presented a model with an undercooled fluid phase that can display both processes evolving on two and on  three well separated time-scales depending on the region of phase diagram analyzed and allows for continuous interpolation between these two extreme situations.
The model describes qualitatively quite well the phenomenology  of viscous liquids in presence of secondary $\beta$ processes. In particular, susceptibility loss shows a distinct
secondary peak when the system is near the region of double time-scale bifurcation (i.e., near the tricritical point across which the 2RSB glass can be reached). This signal is smeared going far from the tricritical point, changing into an excess wing shape, as it has been by schematic MCTs so far, cf. e.g.,  \cite{Goetze02,Goetze04}, based on the Sj{\"o}gren model \cite{Sjogren86,Goetze89b}.

The solution of  thermodynamics of the model below the dynamic transition consists of a hierarchical nesting of processes evolving  on completely different time-scales \cite{Crisanti07,Leuzzi08b}. Such property hints that, even though
taking place on separated time-scales, fast
processes have a relevant influence on slow processes also near $T_d$ from above. This
observation naturally leads to  a comparison with Ngai's Coupling
Model \cite{Ngai03,Ngai04} and stimulates a reflection on the way slow processes dynamics combine into  a  stretched exponential, or, similarly, a Cole-Davidson representation of the 
relaxation in glassy systems.

In our disordered $s$-$p$-spherical spin model  we find that $\alpha$ and $\beta$ relaxation processes do stay apart, cf. Fig. \ref{fig:taus2}, 
down to the dynamic transition, whose counterpart in realistic glass formers is the crossover temperature where the separation of time-scales begins to {\em accelerate}  \cite{Donth01,LeuzziBook}. The relaxation time $\tau_\beta$, actually, increases of several order of magnitude but $\tau_\alpha$ also increases and accelerates faster than $\tau_\beta$. This is apparently in contrast with the observation of Stevenson and Wolynes \cite{Stevenson10}  that approaching $T_d$ (the finite-dimensional analogue of $T_d$, to be precise) $\beta$ process becomes the dominant mode in structural relaxation and the cooperativeness of $\alpha$ and $\beta$ is not distinguishable anymore. Since that result is obtained in the framework of random first order transition systems, to which our model belongs, further investigation is needed to understand the origin of possible substantial differences.

\section*{Acknowledgements}
The authors thank  Simone Capaccioli, Kia Ngai  and Emanuela Zaccarelli for stimulating discussions. 
The research leading to these results has received
 funding from the Italian Ministry of Education,
 University and Research under the Basic Research
 Investigation Fund (FIRB/2008)  program/CINECA grant code RBFR08M3P4.

\appendix\section{Dynamic scaling equation near plateaus}
\label{appendix1}

Let us define    the function $\bar r(q)$ 
\begin{eqnarray}
\bar r(q) &=& \frac{1}{1-q}-\Lambda(q)
\end{eqnarray}
where $\Lambda$ is defined in Eq. (\ref{def:Lambda}),
related to the longest solution $q=\lim_{t\to\infty}C(t)$ of 
(\ref{eq:dyn}) through
\begin{equation}
\bar r(q)=\bar r\; .
\end{equation}
The solution to the above equation corresponds to a minimum of  $\bar r(q)$  ($\bar r^\prime(q)=0$).
Derivatives of $\bar r(q)$ take the form
\begin{eqnarray}
\frac{d^m \bar r(q)}{dq^m} &=& \frac{m !}{(1-q)^{m+1}}-\frac{d^m \Lambda(q)}{dq^m}\, .
\label{eq:der}
\end{eqnarray}

Writing $C(t)=q+\phi(t)$
we can expand $\Lambda[C(t)]$ near $q$, for small $\phi$:
\begin{equation}
\Lambda(q+\phi)=\sum_{m=0}^\infty\frac{\Lambda^{(m)}(q)}{m!}\phi^m
\end{equation}
where 
\begin{equation}
\Lambda^{(m)}\equiv\frac{d^m \Lambda(q)}{dq^m} 
\end{equation}

In Eq. (\ref{eq:dyn}) we can, thus, rewrite the integral
\begin{eqnarray}
&&\int_0^t dt' \Lambda[C(t-t')] \partial_{t'}\phi(t')=
-(1-q)\Lambda(q)  
\label{f:integral}\\
\nonumber
&&\qquad\quad+\sum_{m=1}^\infty \left[
\frac{\Lambda^{m-1}(q)}{(m-1)!}-(1-q)\frac{\Lambda^{m}(q)}{m!}
\right]\phi^m(t)
\\
\nonumber
&&\qquad\quad+\sum_{m=1}^\infty\frac{\Lambda^{m}(q)}{m!}I_m(t)
\\
&&\quad I_m(t)\equiv\int_0^t\, dt^\prime\,\left[ \phi^m(t - t^\prime) - \phi^m (t) \right]\partial_{t^\prime}\phi(t^\prime)
\end{eqnarray}
where $I_m(t)=O(\phi^{m-1}(t))$.

Expanding equation (\ref{eq:dyn}) in powers of $\phi(t)$ and using Eq. (\ref{f:integral}),
after a few algebraic steps we obtain 
\begin{eqnarray}
\nonumber
&&\Gamma_0^{-1}\partial_t \phi(t) + \left[ \bar r + \Lambda(q) - (1-q) \Lambda^{(1)} (q) \right] \phi(t)  \\ \nonumber
&&\quad+ \sum_{m=2}^\infty\left[ \frac{\Lambda^{(m-1)}}{(m-1)!} - (1-q) \frac{\Lambda^{(m)}(q)}{m!} \right]\phi^m(t)   \\ 
&&\quad + \sum_{m=1}^\infty \frac{\Lambda^{(m)}(q)}{m!}I_m(t)= (1-q)\left[ \bar r + \Lambda(q) \right] -1\, .
\label{eq:scale1}
\end{eqnarray}

From (\ref{eq:der}), defining $\gamma_m$ and $\delta_m$ as follows
\begin{eqnarray}
\gamma_m &\equiv& \frac{1}{(1-q)^{m-2}}\\ \nonumber
\delta_m &\equiv& \frac{(1-q)^3}{m!}\frac{d^m}{dq^m}\left[ \bar r(q) - r\right]\; ,
\end{eqnarray}
we can rewrite
\begin{equation}
\frac{\Lambda^{(m)}(q)}{m!}=\frac{1}{(1-q)^3}\left( \gamma_m - \delta_m \right)\, .
\label{eq:lambda}
\end{equation}
Inserting Eq. (\ref{eq:lambda}) in Eq. (\ref{eq:scale1}) we find
\begin{eqnarray} \nonumber
&&\Gamma_0^{-1}\partial_t \phi(t) + \frac{1}{(1-q)^3}\sum_{m=1}\left[ -\delta_{m+1} \right. 
\left. (1-q)\delta_m \right] \phi^m(t) \\ 
&&+\frac{1}{(1-q)^3}\sum_{m=1}\left[ \gamma_m - \delta_m \right] I_m(t) = -\frac{\delta_0}{(1-q)^2}% \\ \nonumber
\end{eqnarray}
that, at the  order $\phi^2$, becomes
\begin{eqnarray}
&&\Gamma_0^{-1}\partial_t \phi(t) + \frac{1}{(1-q)^3} \left[ -\delta_0 + (1-q)\delta_1 \right] \phi(t) 
\label{eq:scale2} \\ \nonumber
&& + \frac{1}{(1-q)^3}\left[ -\delta_1+ (1-q)\delta_2 \right]\phi^2(t) \\ \nonumber  
 &&+ \frac{1}{(1-q)^3}(\gamma_1 - \delta_1) I_1(t)   + o(\phi^3) = -\frac{\delta_0}{(1-q)^2}\,\, .
\end{eqnarray}
If $\bar r (q)$ develops a local minimum we have 
\begin{equation}
\bar r^\prime(q)=\delta_1=0
\end{equation}
near the minimum $\bar r(q)-\bar r \ll 1$ and consequently we define the small quantity 
\begin{equation}
\sigma\equiv \delta_0=(1-q)^3\left[ \bar r(q)-\bar r \right]  \ll 1\, .
\end{equation}
Defining the quantity
\begin{equation}
\overline{m} \equiv \frac{(1-q)^3}{2}\Lambda''(q)
\end{equation}
we can, further, write
\begin{equation}
\delta_2=1-\overline{m}\, .
\end{equation}
We can rewrite Eq. (\ref{eq:scale2}) as follows
\begin{eqnarray} 
&&\Gamma_0^{-1}\partial_t \phi(t) - \frac{\sigma}{(1-q)^3} \phi(t) 
\label{eq:scale3}
 \\ 
\nonumber
&&+\frac{1}{(1-q)^2}\left[ (1-\overline{m})\phi^2 (t)+ I_1(t)\right] + o(\phi^3)= -\frac{\sigma}{(1-q)^2} 
\end{eqnarray}
when $\sigma\to 0$ the solution of Eq. (\ref{eq:scale3}) is of the form
\begin{equation}
\phi(t)= \sigma^{1/2} g(\tau)\, , \;\; \tau=t/t_\sigma=o(1)
\end{equation}
where $g(\tau)$ is solution of the following scaling equation:
\begin{equation}
\nonumber
(1-\overline{m}) g^2(t)+ \int_0^\tau d\tau' \left[ g(\tau-\tau') - g(\tau) \right] \partial_\tau g(\tau')=-1.
\label{eq:rescaled1}
\end{equation}

If the dynamic equation develops a
solution with two plateaus at the values
$q_\kappa$ (with $\kappa=1,2$), we can
fix two rescaled time scales $\tau_\kappa$ where
$t/t_{\sigma_\kappa}=o(1)$ and we
can generalize Eq. (\ref{eq:rescaled1}) to Eq. (\ref{eq:rescaled})
\begin{equation}
\nonumber
(1-\overline{m_\kappa}) g_\kappa^2(t)+ \int_0^\tau d\tau' \left[ g_\kappa(\tau_\kappa-\tau') - g_\kappa(\tau_\kappa) \right] \partial_{\tau'} g_\kappa(\tau')=-1.
\end{equation}

\bibliographystyle{epj}
\bibliography{Lucabib}

\end{document}